\documentclass[aps,prx,reprint,superscriptaddress,amssymb,amsmath,longbibliography]{revtex4-1}
\usepackage{graphicx}
\usepackage{dcolumn}
\usepackage{bm, color}
\usepackage[none]{hyphenat} 
\usepackage{bbm}
\usepackage{braket}

\begin{document}

\title{Dirac nodal lines and induced spin Hall effect in metallic rutile oxides}

\author{Yan Sun}
\affiliation{Max Planck Institute for Chemical Physics of Solids, 01187 Dresden, Germany}

\author{Yang Zhang}
\affiliation{Max Planck Institute for Chemical Physics of Solids, 01187 Dresden, Germany}
\affiliation{Leibniz Institute for Solid State and Materials Research, 01069 Dresden, Germany}

\author{Chao-Xing Liu}
\affiliation{Department of Physics, the Pennsylvania State University, University Park, Pennsylvania 16802-6300, USA}

\author{Claudia Felser}
\affiliation{Max Planck Institute for Chemical Physics of Solids, 01187 Dresden, Germany}

\author{Binghai Yan}
\email{yan@cpfs.mpg.de}
\affiliation{Max Planck Institute for Chemical Physics of Solids, 01187 Dresden, Germany}
\affiliation{Max Planck Institute for the Physics of Complex Systems, 01187 Dresden, Germany}

\begin{abstract}
We have found Dirac nodal lines (DNLs) in the band structures of metallic rutile oxides IrO$_2$, OsO$_2$,
and RuO$_2$
and revealed a large spin Hall conductivity contributed by these nodal lines, 
which explains a strong spin Hall effect (SHE) of IrO$_2$ discovered recently. 
Two types of DNLs exist. The first type forms DNL networks that extend in 
the whole Brillouin zone and appears only in the absence of spin-orbit coupling (SOC), which 
induces surface states on the boundary. Because of SOC-induced band 
anti-crossing, a large intrinsic SHE can be realized in these 
compounds. The second type appears at the Brillouin zone edges and is 
stable against SOC because of the protection of nonsymmorphic symmetry.
Besides reporting new DNL materials, our work reveals the general 
relationship between DNLs and the SHE, indicating a way to apply Dirac nodal materials 
for spintronics.
\end{abstract}

\maketitle

\section{Introduction}
Topological semimetals are an emerging topological phase that has attracted great attention of the condensed-matter community in recent years. Their conduction and valence bands cross each other through robust nodal points or nodal lines in the momentum space~\cite{Wan2011,volovik2003universe,Burkov2011}. Distinct from normal semimetals or metals, resultant Fermi surfaces can be characterized by nontrivial topological numbers, giving rise to exotic quantum phenomena such as Fermi arcs on the surface~\cite{Wan2011}, chiral anomaly in the bulk~\cite{Nielsen1983,Son2013}, anomalous Hall effect (AHE)~\cite{Yang2011QHE,Burkov2014}, and spin Hall effect (SHE)~\cite{Sun2016}.

Dirac and Weyl semimetals are typical three-dimensional topological nodal 
point semimetals, where nodal points and surface Fermi arcs have been 
recently discovered in real materials~\cite{Yan2016}, for example, 
Na$_3$Bi~\cite{Wang2012,Liu2014Na3Bi,Xu2015}, 
TaAs~\cite{Weng2015,Huang2015,Lv2015TaAs,Xu2015TaAs,Yang2015TaAs}, 
and MoTe$_2$~\cite{Soluyanov2015WTe2,Sun2015MoTe2,Deng2016,Jiang2016,Huang2016,Tamai2016}.
Beyond Dirac and Weyl points, new-type symmetry-protected nodal points 
with three-, six- and eight-band crossings in nonsymmorphic space 
groups~\cite{Wieder2016,Bradlyn2016} and nodal points with triple degeneracy 
in symmorphic space groups~\cite{Zhu2016,Winkler2016,Weng2016TaN,Weng2016} 
have been proposed with potential material candidates, such as 
MoP~\cite{Zhu2016,Lv2016}. Furthermore, nodal lines~\cite[and references 
therein]{Burkov2011,Chiu2014,Zeng2015,Fang2015,Chiu2016} can exist 
in two classes of systems, according to the absence and presence of spin-orbit coupling 
(SOC)~\cite{Fang2015,Fang2016}. The first type without SOC has been reported 
in many systems (e.g., LaN) ~\cite{Zeng2015,Weng2015nodeline,Xie2015,Kim2015,Yu2015,Chen2015nano,Chan2016,Li2016,Hirayama2016}. 
The inclusion of SOC will either gap or split the nodal 
lines~\cite{Weng2015,Zeng2015}. The second type requires the protection 
of additional symmetries such as nonsymmorphic 
symmetries~\cite{Fang2015,Young2015,Schoop2016,Bzdusek2016} and 
mirror reflection~\cite{Chiu2014,Weng2015graphene,Ali2014dirac,Bian2016}. 
Nodal lines can cross the whole Brillouin zone (BZ) in a line shape, form 
closed rings inside the BZ, or form a chain containing connected
rings~\cite{Bzdusek2016}. The topological nature of a nodal line can be 
characterized by a quantized Berry phase $\pi$ along a Wilson loop enclosing 
the nodal line~\cite{Burkov2011,Yu2011,Soluyanov2011,Fang2012,Chan2016}. 
On the surface,  nodal line materials were predicted to host drumhead-like 
surface states~\cite{Weng2015graphene}.  Theoretical 
search for exotic nodal phases and corresponding materials launches a race for 
discovering novel topological states in experiments.

Since topological nodal semimetals or metals commonly exhibit nontrivial Berry 
phases and strong SOC, they are expected to reveal a strong SHE as an intrinsic 
effect from the band structure~\cite{Sun2016}. The intrinsic SHE, in which the 
charge current generates the transverse spin current, is intimately related with 
the Berry phase and SOC~\cite{Sinova2015}. Four of us recently found that the 
first type of Dirac nodal lines (DNLs) (without SOC) can induce a strong intrinsic 
SHE when turning on SOC, for example, in the TaAs-family Weyl semimetals~\cite{Sun2016}. 
Furthermore, it provokes us to search for topological nodal systems among known 
SHE (or AHE) materials. Therefore, our attention has been drawn to SHE material 
IrO$_2$ discovered recently~\cite{Fujiwara2013}, where its thin films act as 
efficient spin detectors~\cite{Fujiwara2013,Qiu2015} via the inverse SHE that 
converts the spin current to the electric voltage. However, the microscopic understanding 
of the SHE is still missing for this metallic oxide. Furthermore, the strong SOC of 
the Ir-$5d$ orbitals and nonsymmorphic symmetries of its rutile crystal structure 
imply that IrO$_2$ may host topological nodes in the band structure. 
Additionally, this oxide has been used for electrodes in various applications such as 
catalysts in water splitting~\cite[and references therein]{Tilley2010} and 
ferroelectric memories for a long time~\cite{Scott2000}.

In this article, we have theoretically investigated the topology of the band structure 
of metallic rutile oxide IrO$_2$ and similar oxides RuO$_2$ and OsO$_2$. We observe two types of DNLs in their band structures. The first type extends the whole BZ and forms a square-like DNL network in the absence of SOC, resulting in surface states. Joint points of the network are six- and eight-band-crossing points at the center and boundary of the BZ, respectively. These DNLs become gapped and lead to a strong SHE when SOC exists. The second type is stable against SOC and appears at the edges of the tetragonal BZ, which is protected by nonsymmorphic symmetries.

\section{Methods}

To investigate the band structure and the intrinsic SHE, we have performed $ab~initio$ 
calculations based on the density-functional theory (DFT)  with the localized 
atomic orbital basis and the full potential as implemented in the 
full-potential local-orbital (FPLO) code~\cite{Koepernik1999}. The exchange-correlation 
functionals were considered at the generalized gradient approximation (GGA)
level~\cite{perdew1996}. We adopted the experimentally measured lattice structures for 
$R$O$_2$ ($R$ = Ir, Os, and Ru) compounds.
By projecting the Bloch states into a highly symmetric atomic orbital like 
Wannier functions (O-$p$ and $R$-$d$ orbitals), we constructed tight-binding Hamiltonians and 
computed the intrinsic spin Hall conductivity (SHC) by the linear-response Kubo formula approach in
the clean limit~\cite{Sinova2015, Xiao2010},
\begin{equation}
\begin{aligned}
\sigma_{ij}^{k}=e\hbar\int_{_{BZ}}\frac{d\vec{k}}{(2\pi)^{3}}\underset{n}{\sum}f_{n\vec{k}}\Omega_{n,ij}^{k}(\vec{k}), \\
\Omega_{n,ij}^{k}(\vec{k})=-2Im\underset{n'\neq n}{\sum}\frac{<n\vec{k}| J_{i}^{k}|n'\vec{k}><n'\vec{k}| v_{j}|n\vec{k}>}{(E_{n\vec{k}}-E_{n'\vec{k}})^{2}}
\end{aligned}
\label{SHC}
\end{equation}
where $f_{n\vec{k}}$ is the Fermi--Dirac distribution for the $n$-th band.
The spin current operator is $J_{i}^{k}=\frac{1}{2}\left\{ \begin{array}{cc}
{v_{i}}, & {s_{k}}\end{array}\right\} $, with the spin operator ${s}$,
the velocity operator
${v_{i}}=\frac{1}{\hbar}\frac{\partial {H}}{\partial k_{i}}$,
and $i,j,k=x,y,z$. $| n\vec{k}>$ is the eigenvector for the Hamiltonian
${H}$ at the eigenvalue $E_{n\vec{k}}$. $\Omega_{n, ij}^{k}(\vec{k})$
is referred as the spin Berry curvature as analogy to the ordinary Berry curvature.
A $500 \times 500 \times 500$ $k$-grid in the BZ was used for the integral of the SHC.
The SHC $\sigma_{ij}^{k}$ refers to the spin current ($j_i^{s,k}$), which flows 
along the $i$-th direction with the spin polarization along $k$,
generated by an electric field ($E_j$) along
the $j$-th direction, i.e., $j_i^{s,k} = \sigma_{ij}^{k} E_j$.

\section{Results and Discussions}

\begin{figure}[htbp]
\begin{center}
\includegraphics[width=0.5\textwidth]{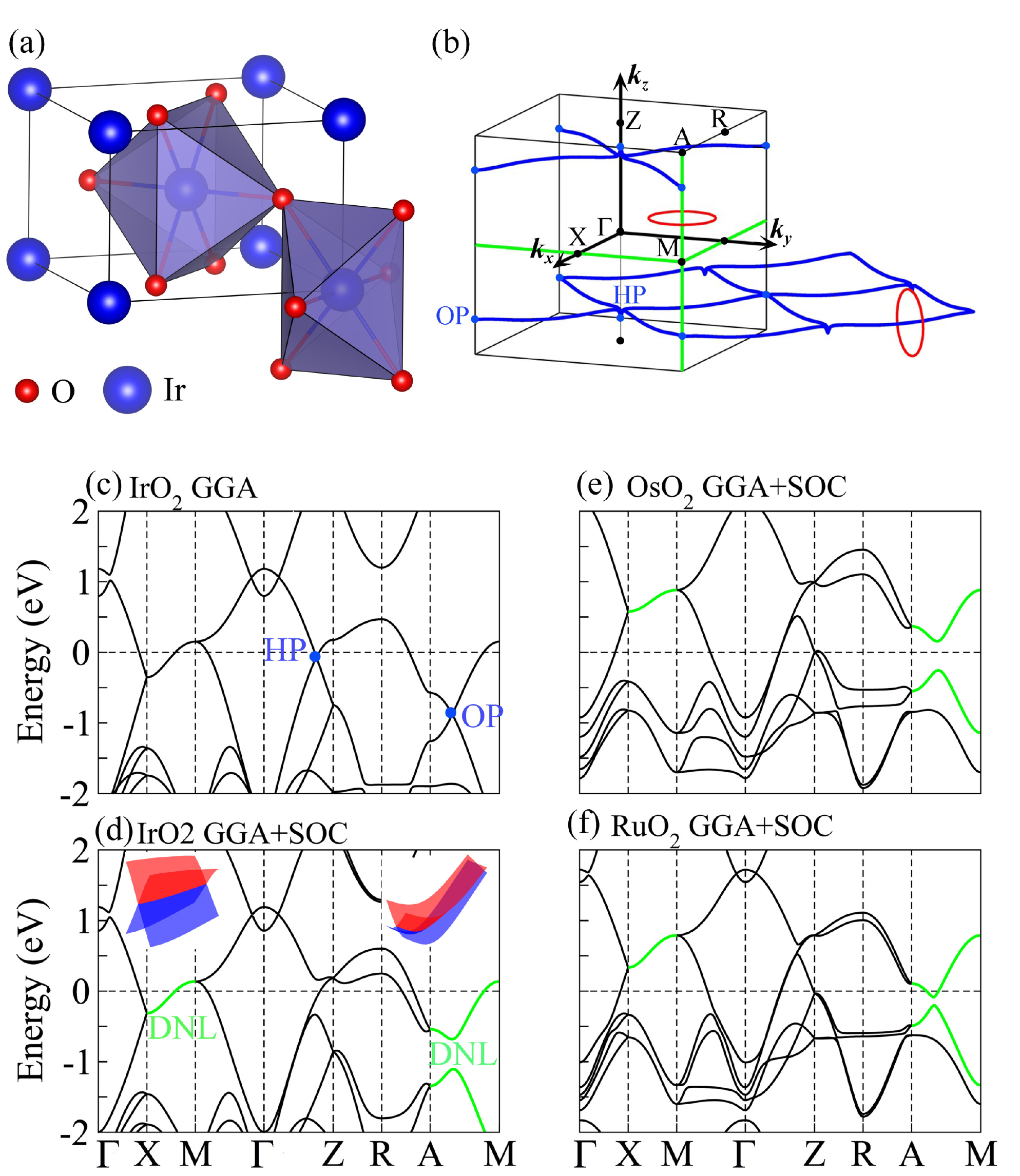}
\end{center}
\caption{
(a) Rutile lattice structure of IrO$_2$. 
(b) The tetragonal BZ of IrO$_2$. 
DNLs (blue lines) without SOC form 
networks inside the BZ, and the corresponding Wilson loop around it
(red circle) gives a $\pi$ Berry phase. The connection points 
of the network are hexatruple points (HPs) and octuple points (OPs). 
The DNLs (green) with SOC are lines along $X$-$M$ and $M$-$A$ at the BZ boundary,
and the Wilson loop around them has zero Berry phase.
(c) Band structure for IrO$_2$ without SOC. The HP and OP are indicated by blue dots.
(d-f) Band structures for IrO$_2$, OsO$_2$, and RuO$_2$ with SOC. The DNLs
on $X-M$ and $M-A$ are highlighted by green lines. In (d), the 3D band 
dispersion around the DNLs is shown as insets.
}
\label{lattice}
\end{figure}

\subsection{Nonsymmorphic symmetry}
Three compounds $R$O$_2$ ($R$ = Ir, Os, and Ru) share the rutile-type 
lattice structure with space group $P4_2/mnm$ (No. 136), as shown in \textbf{Fig. 1}.
A primitive unit cell contains two $R$ atoms that sit at the corner and 
center of the body-centered tetragonal lattice, respectively.
One $R$ atom is surrounded by six O atoms that form a distorted octahedron.
For space group No. 136, we have the following generator operations,
\begin{equation*}
E, m_z, P, n_x \equiv \{ m_x | \vec{\tau} \}, n_{4z} \equiv \{ c_{4z} | \vec{\tau} \}, 
\end{equation*}
where $m_{x,z}$ are mirror reflections, $c_{4z}$ is the four-fold rotation, $P$ is the inversion symmetry, $n_x$ and $n_{4z}$ represent nonsymmorphic symmetries, and $\vec{\tau}=(\frac{1}{2},\frac{1}{2},\frac{1}{2})$ is the translation of one-half of a body diagonal. Additionally; the time-reversal symmetry $T$ also appears for $R$O$_2$ systems.

It is known that electronic bands are doubly degenerate (considering spin and SOC) 
at every $k$-point of the BZ owing to the coexistence of $P$ and $T$. Furthermore, a generic 
nonsymmorphic symmetry leads to new band crossings and thus higher degeneracies 
at the BZ boundary. Therefore, the coexistence of $P$, $T$, and nonsymmorphic symmetries 
guarantees four-fold or even larger degeneracies at some $k$-points of the BZ boundary. 
As discussed in the following, IrO$_2$ exhibits a four-fold degeneracy at the BZ edge lines, 
$X(Y)-M$ and $M-A$ (also see \textbf{Fig. 1d}). Consequently, one requires $4n$ electrons 
for filling these bands to obtain a band insulator. However, a primitive unit cell contains 
two IrO$_2$ formula units and in total 42 valence electrons, failing to satisfy the 
precondition of a band insulator in this nonsymmorphic space 
group~\cite{Parameswaran2013,Bradlyn2016}. Therefore, IrO$_2$ is constrained by the lattice 
symmetry to be a band metal in the weak interaction case. Recently, it has been reported that IrO$_2$ 
cannot become a Mott insulator because of a large bandwidth~\cite{Kahk2014,Kim2016IrO2}, while 
several seemingly similar iridates (e.g., Sr$_2$IrO$_4$) are known as $5d$ Mott insulators 
with the $J_{eff}=1/2$ state~\cite{Kim2008,Kim2009}.

\begin{figure*}[htbp]
\begin{center}
\includegraphics[width=0.85\textwidth]{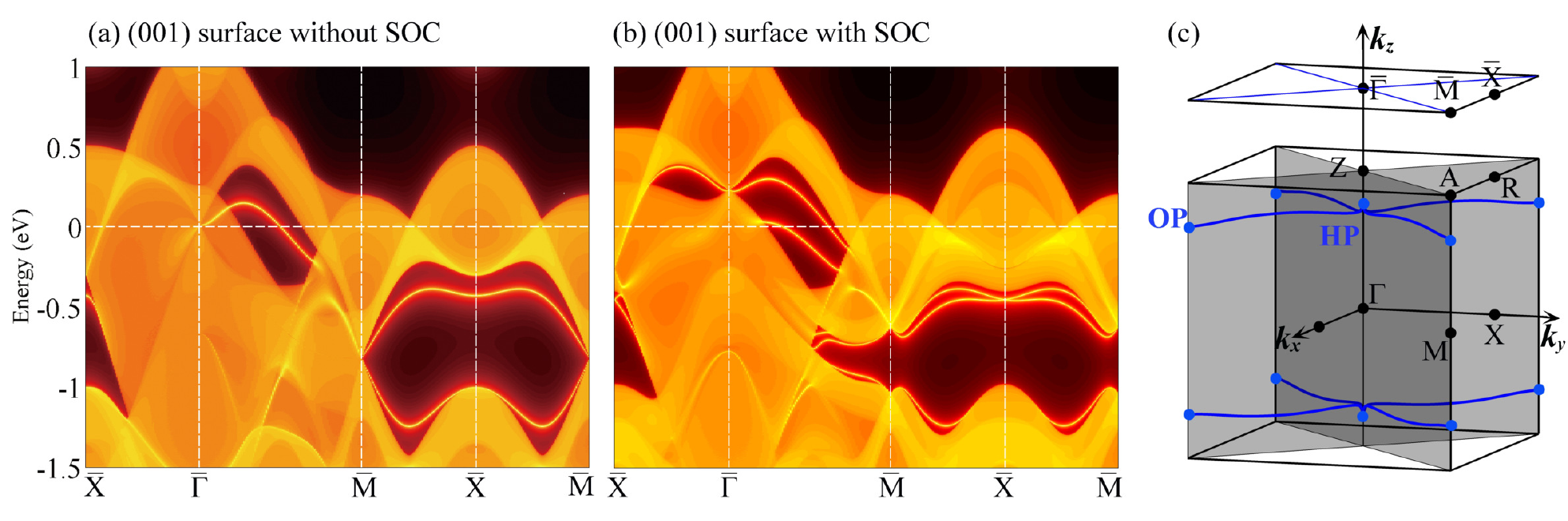}
\end{center}
\caption{
Surface band structure for IrO$_2$ projected to the (001) surface without (a) and with (b) SOC. The brightness of color represents the weight of surface states. (c) The projection of the bulk BZ to the surface BZ with DNLs.}
\label{surface}
\end{figure*}

\subsection{Dirac Nodal lines without SOC}

We first investigate the band structures without including SOC. 
As shown in \textbf{Fig. 1c}, six-band and eight-band crossing points 
(including spin) appear at the $\Gamma-Z$ and $M-A$ axes, respectively, 
which are noted as a hexatruple point (HP) and an octuple point (OP), respectively. 
There is a DNL connecting neighboring HP and OP in the BZ, forming a network 
in the $k$-space, as indicated by \textbf{Figs. 1b } and \textbf{3a}. 
Two layers of networks are present above and below the $k_z=0$ plane respectively, 
and can be transformed to each other by $P$ or $T$. DNLs exist inside the (110) and ($\bar{1}$10) mirror planes and originate from crossing of $d_{x^2-y^2}$ and $d_{xz,yz}$ bands. Here, $d_{x^2-y^2}$ and $d_{xz,yz}$ bands are all doubly degenerate owing to $P$ and $T$ and exhibit opposite eigenvalues $-1$ and $+1$, respectively, for each mirror reflection. The mirror symmetry protects the four-band-crossing in the absence of SOC.

The topology of a DNL is characterized by the nontrivial Berry phase (or winding number) along a closed path that includes the DNL. We choose a loop, along which the system is fully gapped as indicated in \textbf{Fig. 1b}~\cite{Yu2011,Soluyanov2011}. The Berry phase for all ``occupied'' bands is found to be a quantized value, $\pi$. This nonzero Berry phase further leads to surface states. When projecting to the (001) surface, the nodal band structure exhibits a local energy gap between $\bar{M}$ and $\bar{X}$ in the surface BZ. Two layers of DNL networks overlap in the (001) surface projection. Thus, one can observe two sets of surface states (spinless) connecting OPs between two adjacent $\bar{M}$ points inside the gap in \textbf{Fig. 2a}. 

When SOC is included and the $SU(2)$ symmetry is broken, these DNLs including HPs and OPs are gapped (see \textbf{Fig. 1d}), since there is no additional symmetry protection. On the surface, original spinless surface states split into two Rashba-like spin channels (see \textbf{Fig. 2b}). Because of the lack of robust symmetry protection (e.g., $P$ is commonly breaking on the surface), these surface states may appear or disappear according to the surface boundary condition.

\subsection{Dirac nodal lines with SOC}

The $k_x=\pi$ and $k_y=\pi$ planes are actually Dirac nodal planes in the absence of SOC. 
The presence of SOC gaps these planes and only leaves DNLs along some high-symmetry lines, 
$X$-$M$ and $M$-$A$. By taking the DNL along $X$-$M$ as an example, we can understand 
the four-fold degeneracy by 
considering time reversal symmetry $T$, point group symmetries, $m_z$, $P$, and nonsymmorphic 
symmetry $n_x$. Since $[m_z, H]=0$ in the $k_z=0$ plane, we can choose the eigenstates 
of Hamiltonian $H$ with definite mirror parity for $\vec{k}$ along $X$-$M$,
\begin{equation}
\begin{aligned}
H(\vec{k})|\phi_{\pm}(\vec{k})>&=E_{\pm}(\vec{k})|\phi_{\pm}(\vec{k})> \\
m_{z}|\phi_{\pm}(\vec{k})>&=\pm i|\phi_{\pm}(\vec{k})>
\end{aligned}
\end{equation}
where $i$ is from the spin. First, we know $[TP, m_z]=0$. Therefore, we have 
\begin{equation}
\begin{aligned}
m_{z} (TP|\phi_{\alpha}(\vec{k}))> = TP (m_z |\phi_{\alpha}(\vec{k})>) \\
= TP (i\alpha |\phi_{\alpha}(\vec{k})>) = -i \alpha (TP |\phi_{\alpha}(\vec{k})>)
\end{aligned}
\end{equation}
where $\alpha=\pm 1$. Therefore, $TP |\phi_{\alpha}(\vec{k})>$ is also an eigenstate 
at $\vec{k} ~(TP\vec{k}=\vec{k})$ with the mirror parity $-i\alpha$, where $-$ sign is 
from the complex conjugate in $T$. Next, we consider the commutation between $m_z$ and 
the glide mirror symmetry $n_x$, both of which act in real space $(x,y,z)$ and the spin 
space simultaneously. In real space, we have 
\begin{equation}
\begin{aligned}
(x,y,z) &\xrightarrow{n_x} (-x+\frac{1}{2},y+\frac{1}{2},z+\frac{1}{2}) \\
          &\xrightarrow{m_z} (-x+\frac{1}{2},y+\frac{1}{2},-z-\frac{1}{2}) \\
(x,y,z) &\xrightarrow{m_z} (x,y,-z) \\
          &\xrightarrow{n_x} (-x+\frac{1}{2},y+\frac{1}{2},-z+\frac{1}{2}) \\       
m_z n_x &= T_{(0,0,-1)} n_x m_z \\
\end{aligned}
\end{equation}
where $T_{(0,0,-1)} = e^{i k_z} $ is the translation operator when acting on the 
Bloch wavefunction. In the spin space, $m_z = i \sigma_z$ and $n_x = i \sigma_x$ 
and thus $m_z n_x = - n_x m_z$. By combining the real space and the spin space, we obtain
\begin{equation}
\begin{aligned}
 m_z n_x = e^{i k_z} (-1) n_x m_z = -n_x m_z\\
\end{aligned}
\end{equation}
Therefore, $n_x |\phi_{\alpha}(\vec{k})>$ is also an eigenstate at $\vec{k}~(n_x \vec{k}=\vec{k})$, but with a mirror parity $-i \alpha$.
Further, the combination of $TPn_x$ leads to one more state $TPn_x |\phi_{\alpha}(\vec{k})>$ with a mirror parity $+i \alpha$.

In total, we can have four eigenstates, $|\phi_{\alpha}(\vec{k})>$, $TP|\phi_{\alpha}(\vec{k})>$, $n_{x}|\phi_{\alpha}(\vec{k})>$, and $TPn_{x}|\phi_{\alpha}(\vec{k})>$ for $\vec{k}~(\pi,k_{y},0)$ along $X$-$M$. The mirror parities of $m_z$ are $+i \alpha$ for $|\phi_{\alpha}(\vec{k})>$ and $TPn_{x}|\phi_{\alpha}(\vec{k})>$, and $-i \alpha$ for $TP|\phi_{\alpha}(\vec{k})>$ and $n_{x}|\phi_{\alpha}(\vec{k})>$. Next, we will prove that they are orthogonal to each other, i.e., two eigenstates with the same mirror parity are orthogonal, $<\phi_{\alpha}(\vec{k})|TPn_x |\phi_{\alpha}(\vec{k})>=0$. This requires that $TPn_x$ is anti-unitary. In real space, we have 
\begin{equation}
\begin{aligned}
(x,y,z) &\xrightarrow{Pn_x} (x-\frac{1}{2},-y-\frac{1}{2},-z-\frac{1}{2}) \\
          &\xrightarrow{Pn_x} (x-1,y,z) \\
\end{aligned}
\end{equation}
In the spin space, $P=1$ and $n_x=i\sigma_x$. Therefore, we have $(Pn_x)^2 = - e^{i k_x}$.
Considering $T^2 = -1$ for spinful fermions, we obtain 
$(TPn_x)^2 = e^{i k_x} = -1$ for $\vec{k}=(\pi, k_y, 0)$. Since $TPn_x$ is an 
anti-unitary operator that satisfies $<\phi|\psi>=<TPn_x\psi | TPn_x \phi>$, we have
\begin{equation}
\begin{aligned}
& <\phi_{\alpha}(\vec{k})|TPn_x |\phi_{\alpha}(\vec{k})>\\
&=<(TPn_x)^2\phi_{\alpha}(\vec{k})|TPn_x |\phi_{\alpha}(\vec{k})> \\
&=-<\phi_{\alpha}(\vec{k})|TPn_x |\phi_{\alpha}(\vec{k})>=0
\end{aligned}
\end{equation}
which means the states $TP|\phi_{\alpha}(\vec{k})>$ and $n_x|\phi_{\alpha}(\vec{k})>$
are orthogonal to each other.

Therefore, we prove that there are four degenerate orthogonal eigenstates along the $X$-$M$ line. 
We point out that the nonsymmorphic symmetry is crucial to protect the four-fold degeneracy, while only $P, T$, and $m_z$ cannot stabilize DNLs.
Likewise, we can also understand the four-fold degeneracy along the $M$-$A$ line considering $m_z$, $P$, $T$, and another nonsymmorphic symmetry $n_{4z}$.

These DNLs protected by nonsymmorphic symmetries are similar to those DNLs observed at the BZ edges in ZrSiS and HfSiS (e.g.,~\cite{Schoop2016,Chen2017}). 
We find that the Berry phase along a loop (indicated in \textbf{Fig. 1b}) including such a DNL is zero. Therefore, we do not expect apparent topological surface states related to these DNLs. 
It is still interesting to point out that t he density of states scales linearly to the energy for DNLs. 
Then one can expect different correlation effects in DNL semimetals from a nodal point semimetal and a normal metal~\cite{Huh2016,Fang2015}. 
For IrO$_2$, such DNLs appear at the Fermi energy, while they stay far below the Fermi energy in ZrSiS-type compounds. These DNLs may be responsible for the high conductivity and large magneroresistance~\cite{Lin2004,Ryden1972}.

\subsection{Dirac nodal lines and spin Hall effect}

The first type of DNLs, DNL networks without SOC, indicate the existence of a strong SHE in $R$O$_2$. 
It is known that band anti-crossing induced by SOC can lead to a large intrinsic SHC. 
To maximize the SHC, one needs to increase the number of band anti-crossing points, i.e., nodal points in the absence of SOC.
Therefore, a DNL, an assemble of continual nodal points in the BZ, can induce a strong SHE. 
The DNL networks that constitute many DNLs will further enhance the SHE. 
Because such type of DNLs without SOC are usually protected by the mirror symmetry, 
it will be insightful to look for SHE materials in space groups that host many mirror or glide mirror planes.
This argument seems consistent with the fact that current best SHE materials are usually Pt and W 
metals(e.g.~\cite{Tanaka2008,Guo2008}) from the high-symmetry space groups.

\begin{table}[t]
  \centering
  \caption{SHC for IrO$_2$, OsO$_2$, and RuO$_2$. The Fermi energy is set to
the charge neutral point. The SHC is in units
of $(\hbar/e) (\Omega\cdot cm)^{-1}$.}
    \label{tab:table1}
  \begin{tabular}{rrrrr}
  \hline
  \hline
                     &     IrO$_2$  &     OsO$_2$  &     RuO$_2$    \\
  \hline
  $\sigma_{xy}^{z}$  &        8  &       9  &       83    \\
  $\sigma_{zx}^{y}$  &    --253  &    --311  &    --238    \\
  $\sigma_{yz}^{x}$  &    --161  &    --541  &    --284    \\
\hline
\hline
  \end{tabular}
\end{table}

\begin{figure}[htbp]
\begin{center}
\includegraphics[width=0.45\textwidth]{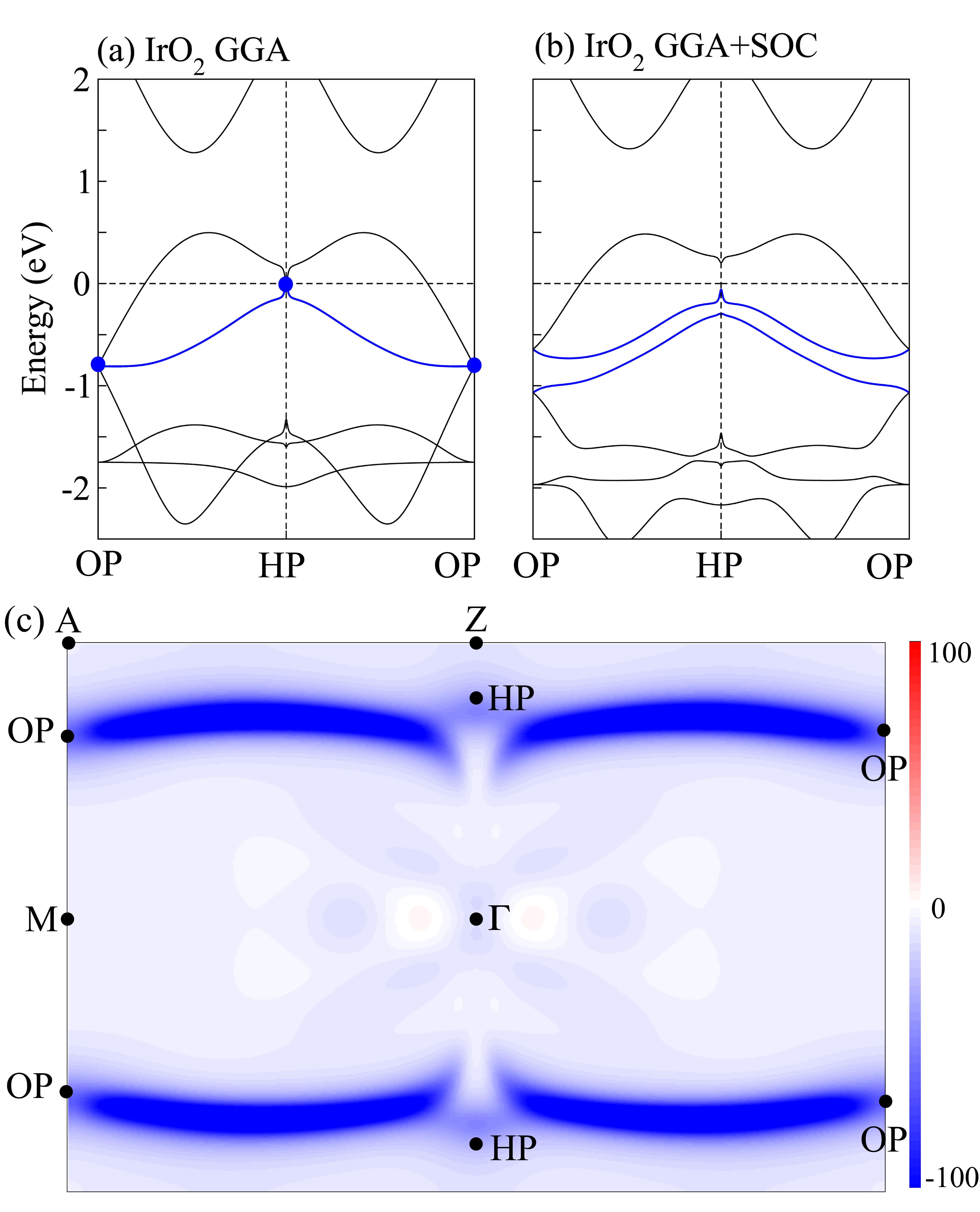}
\end{center}
\caption{
DNLs and SHE in IrO$_2$.
(a) Band structure without SOC along a first-type 
DNL. Blue curves represent the DNL that connects the HP and the 
OP in the (110) or ($\bar{1}$10) mirror plane.
(b) Band structure with SOC along the same line. 
A gap opens along the nodal line due to SOC.
(c) Distribution of the spin Berry curvature 
$\Omega^x_{n,yz}(\vec{k})$ inside the (110) mirror 
plane. Here, $\Omega^x_{n,yz}(\vec{k})$ is summarized 
over all bands below the SOC split gap. It is clear that 
the spin Berry curvature is dominantly contributed by the 
DNL regions. The blue and red colors represent 
negative and positive values of the spin Berry 
curvature in unit of $(\AA^{-1})^{2}$
}
\label{local}
\end{figure}

Next, we compute the intrinsic SHC for $R$O$_2$ compounds following Eq.~\ref{SHC} 
based on the band structure including SOC. The SHC $\sigma_{ij}^{k}$ is a 
second-order tensor with 27 elements. 
The number of independent nonzero elements of the SHC tensor is constrained
by the symmetry of the system.
We have only three independent nonzero elements, 
$\sigma_{xy}^{z}=-\sigma_{yx}^{z}$, $\sigma_{zx}^{y}=-\sigma_{zy}^{x}$, and $\sigma_{yz}^{x}=-\sigma_{xz}^{y}$.
We list their values in Table I for all the three compounds.
The SHC of OsO$_2$ is larger than that of RuO$_2$, and Os-$5d$ bands exhibit much stronger SOC than Ru-$4d$ bands.
The SHC is dependent on the Fermi energy of the system. 
Because Ir has one more electron than Os/Ru, the Fermi surface of IrO$_2$ is higher than that of OsO$_2$ and RuO$_2$ although their band structures look very similar (see \textbf{Figs. 1d-1f}).
Thus,  IrO$_2$ displays a smaller SHC than OsO$_2$ although Ir has stronger SOC.
For all the three compounds, the SHC is very anisotropic, as can be seen from Table I.
Here, the largest SHC of $R$O$_2$ is still smaller than that of Pt [$\sigma^z_{xy} \sim 2000 (\hbar/e) (\Omega\cdot cm)^{-1}$]. 
However, $R$O$_2$ may exhibit a large spin Hall angle (the ratio of the SHC over the charge conductivity) owing to their high resistivity compared to a pure metal, as already found in IrO$_2$ ~\cite{Fujiwara2013}.

Here, we also demonstrate the direct correspondence between DNLs and the SHC. 
Since the SHC is obtained by integrating the spin Berry curvature over the BZ,
we show the distribution of $\Omega^x_{n,yz}(\vec{k})$ in the (110) mirror plane that hosts DNLs.
In the band structure along DNLs in \textbf{Fig. 3}, the SOC clearly gaps a DNL that connects a HP and an OP.
Correspondingly, one can find two ``hot lines'' of the spin Berry curvature, 
which is the anti-crossing region of the DNLs.
Thus, it is clear that the first-type DNLs contribute to a large SHC for IrO$_2$-type materials. 
In contrast, the second-type DNLs at the BZ edges (e.g., that along $M$-$A$ in \textbf{Fig. 3c}) 
show a slight contribution to the SHC.

\section{Conclusions}

To summarize, we found two types of DNLs in metallic rutile oxides IrO$_2$, 
OsO$_2$, and RuO$_2$.
First-type DNLs form networks that extend in the whole BZ and appear only in the absence of SOC, 
which induces surface states at the boundary. The second type of DNLs is stable against SOC because of the 
protection of nonsymmorphic symmetry. 
Because of the SOC-induced gap for first-type DNLs, a large intrinsic SHE 
can be realized in these compounds. 
This explains the strong SHE observed in IrO$_2$ in the previous experiment. 
Moreover, our calculation suggests that OsO$_2$ will behave even better than IrO$_2$ in SHE devices, 
as OsO$_2$ shows a larger intrinsic SHC. Our work implies that first-type DNLs 
(or nodal points that can be gapped by SOC) and the SHE maybe commonly related to each other, 
indicating new guiding principles to search for DNLs in SHE materials or enhance the SHE 
by DNLs in the band structure. For example, it will be insightful to look for SHE materials 
in high-symmetry space groups with many mirror or glide mirror planes, which can induce the 
first type of DNLs.

\begin{acknowledgments}
We thank Jakub Zelezny and Lukas M\"uchler for fruitful discussions.
This work was financially supported by the ERC 
(Advanced Grant No. 291472 ``Idea Heusler").
Y.Z. and B.Y. acknowledge German Research Foundation (DFG) SFB 1143.
\end{acknowledgments}


\bibliography{TopMater}

\end{document}